\documentclass[conference]{IEEEtran}
\IEEEoverridecommandlockouts
\usepackage{cite}
\usepackage{amsmath,amssymb,amsfonts}
\usepackage{graphicx}
\usepackage{textcomp}
\usepackage{xcolor}
\def\BibTeX{{\rm B\kern-.05em{\sc i\kern-.025em b}\kern-.08em
    T\kern-.1667em\lower.7ex\hbox{E}\kern-.125emX}}
\usepackage{makecell}
\usepackage{amsmath,amssymb,amscd,amsthm,amsfonts}
\usepackage{colortbl}
\usepackage{soul}
\usepackage{multirow}
\usepackage{booktabs}
\usepackage{makecell}
\usepackage{tcolorbox}
\usepackage{fancybox} 
\usepackage{adjustbox} 
\usepackage{anyfontsize}
\usepackage{graphicx}
\usepackage{algorithm}
\usepackage{algpseudocode}
\usepackage{titlesec}
\usepackage[colorlinks=true,linkcolor=blue,anchorcolor=black,citecolor=red,filecolor=black,menucolor=black,runcolor=black,urlcolor=black]{hyperref} 

\begin{document}

\title{AI-Enhanced Inverter Fault and Anomaly Detection System for Distributed Energy Resources in Microgrids
\thanks{S.-R. Kasimalla, K. Park, and J. Hong are with the Department of Electrical and Computer Engineering, University of Michigan -- Dearborn, Dearborn, MI, 48128 USA (emails: sweraka@umich.edu; kuchan@umich.edu, jhwr@umich.edu).\\
Y.-J. Kim is with the Department of Electrical Engineering, Pohang University of Science and Technology, Pohang 37673, South Korea (e-mail: powersys@postech.ac.kr).\\ 

H. Lee is with DTE Energy, Detroit, MI, 48226 USA (e-mail: hyojong.lee@dteenergy.com).
}
}

\author{\IEEEauthorblockN{Swetha Rani Kasimalla, \textit{Graduate Student Member, IEEE}, Kuchan Park, \textit{Graduate Student Member, IEEE},\\ Junho Hong, \textit{Senior Member, IEEE}, Young-Jin Kim, \textit{Senior Member, IEEE}, HyoJong Lee, \textit{Member, IEEE}}}

\maketitle

\begin{abstract}
The integration of Distributed Energy Resources (DERs) into power distribution systems has made microgrids foundational to grid modernization. These DERs, connected through power electronic inverters, create a power electronics-dominated grid architecture, introducing unique challenges for fault detection. While external line faults are widely studied, inverter faults remain a critical yet underexplored issue. This paper proposes various data mining techniques for the effective detection and localization of inverter faults—essential for preventing catastrophic grid failures. Furthermore, the difficulty of differentiating between system anomalies and internal inverter faults within Power Electronics-Driven Grids (PEDGs) is addressed. To enhance grid resilience, this work applies advanced artificial intelligence methods to distinguish anomalies from true internal faults, identifying the specific malfunctioning switch. The proposed FaultNet-ML methodology is validated on a 9-bus system dominated by inverters, illustrating its robustness in a PEDG environment.
\end{abstract}
\titlespacing*{\section}{0pt}{-1pt}{5pt}
\section{Introduction}
Today’s power systems are evolving with a focus on sustainability, efficiency, and resilience, driven largely by the development of microgrids and the advancement of inverter-based resources (IBRs). IBRs bring notable improvements in operational flexibility, reliability, and minimize energy loss; however, their integration introduces challenges across several critical dimensions. Managing diverse sources in PEDG systems, such as solar, wind, and battery storages that requires coordinated control to ensure balanced power distribution and system stability. Furthermore, PEDG systems rely on smart grid integration and extensive real-time data sharing, raising privacy concerns related to sensitive user information. Stability, planning, and resilience are also crucial considerations: integrating sources with varying dynamic behaviors risks fluctuations in voltage and frequency, which can potentially damage equipment. Strategic planning is required to meet growing demand while efficiently allocating resources. Additionally, as digital control becomes more prevalent, PEDG systems are increasingly vulnerable to cyber-attacks that could disrupt operations and impact system functionality.

 Inverter components within PEDG systems, particularly in renewable environments, face high levels of stress, making them susceptible to faults. Generally inverter faults include switch short circuits and open-switch faults. Short circuits arise from overvoltages or abnormal signals, causing large, rapid currents that may lead to open-switch faults if mitigated. Open-switch faults, caused by issues such as bond wire lifting and gate drive failures, distort three-phase currents and voltages, creating harmonic losses and stressing other components, ultimately compromising power quality. This vulnerability underscores the need for rapid, efficient fault detection and localization, especially as smart grid integration exposes inverter control to cyber threats. Recent advancements in big data processing now enable AI-based solutions that can effectively distinguish between internal faults and anomalous operations, making them essential for addressing these complex challenges.

 There has been limited research on inverter fault diagnosis methods, especially within extensive microgrid environments. Previous works, such as those by ~\cite{8590418} and ~\cite{8329516}, have introduced open-circuit fault diagnosis in motor-driven systems, while ~\cite{9544141} investigated open-circuit fault detection in a 7-level hybrid active neutral point clamped (7L-ANPC) multilevel inverter for the first time. More recent studies ~\cite{10214285,10215460,10451077} have made advancements by applying data-driven approaches to diagnose open-switch faults. However, none of these studies have examined fault diagnosis in a fully integrated microgrid framework, where inverters are tested in interconnected settings with other loads and communication channels. 
 In recent studies, various fault detection approaches have been proposed for microgrid systems, each focusing on different fault types, detection mechanisms, and implementation techniques.
 
This paper addresses these gaps by proposing a machine-learning-based method for inverter fault detection and localization within a 9-bus microgrid system, including integrated Battery Energy Storage Systems (BESS) and solar PV units. Data collected from this microgrid configuration was used to train and validate machine learning models, including K-Nearest Neighbors (KNN) and other classification algorithms. This methodology facilitates rapid and accurate detection of inverter faults, enabling timely corrective action to ensure clean sinusoidal output for local and critical loads. The proposed approach can identify single and multiple switch faults in a three-phase three-leg voltage source inverter, as well as detect anomalies like false data injection scenarios—an essential consideration in the context of real-time grid operation. The following key contributions encapsulate the novelty and importance of this research:
\begin{itemize}
    \item The paper introduces a comprehensive approach to classify inverter faults within complex microgrid systems, enabling a detailed understanding of fault types and their operational impacts.
    \item The proposed method differentiates between anomalies and hardware malfunctions, particularly in gate drive units. This distinction is critical for targeted responses, enhancing system security and reliability
    \item Upon identifying a malfunction, the methodology accurately localizes the specific inverter fault, distinguishing between single-switch and multi-switch faults, facilitating efficient troubleshooting and maintenance.
    \item Various data-driven models were evaluated for their performance in fault classification, leading to the selection of the most accurate model. This evaluation ensures the proposed approach is robust and reliable for practical applications.
\end{itemize} 
The remainder of this paper is structured as follows: Section II elaborates on the architecture of the microgrid framework and the fault classification approach.  Section IV contains the simulation results and discussion, and Section V is the conclusion followed by acknowledgment as in section VI.
\section{System Framework and Fault Detection Algorithm}
\subsection{Development of Microgrid Framework}
\begin{figure}
\includegraphics[width=1\columnwidth]{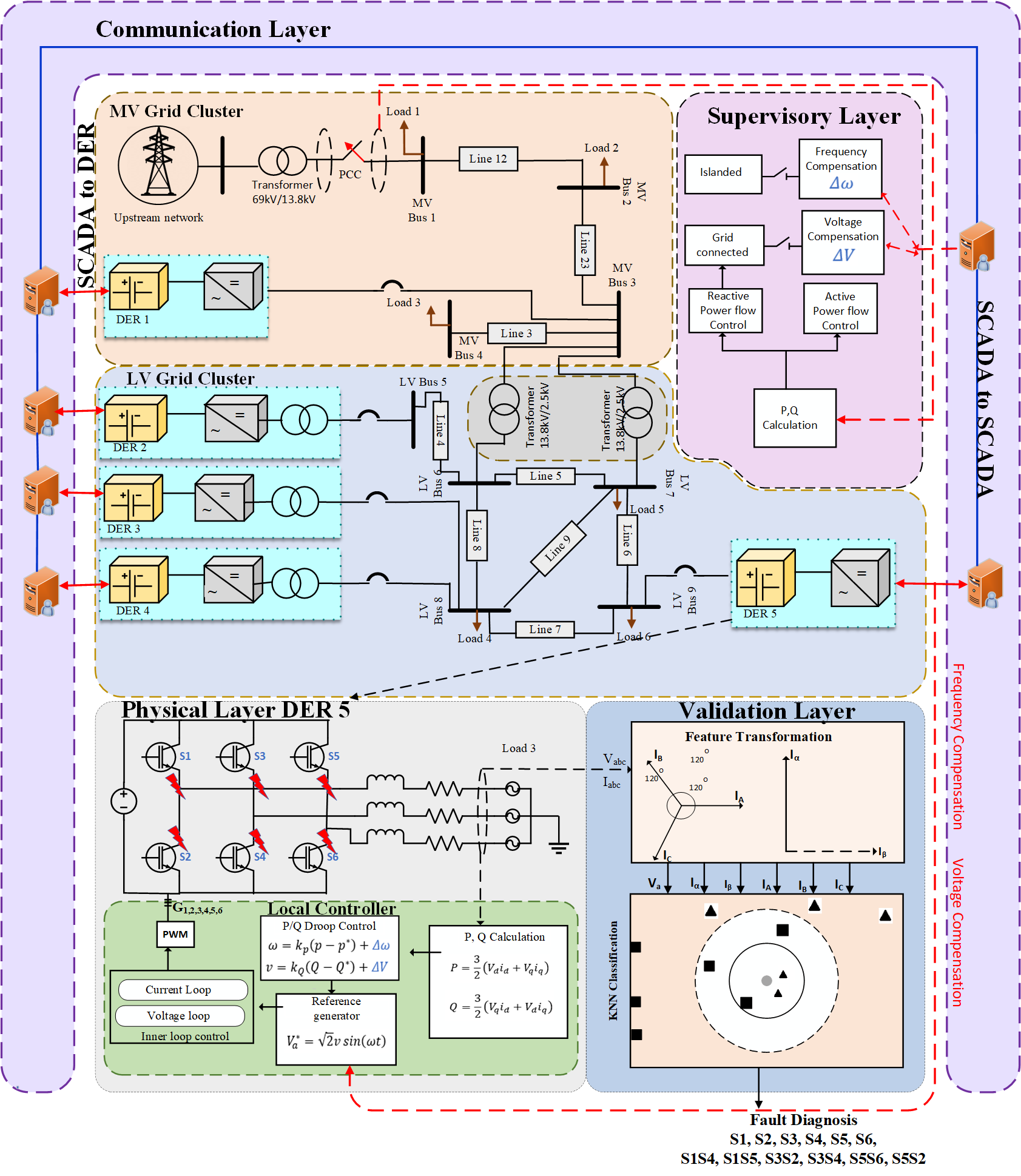}
\caption{Multi-Layer Microgrid Framework: Integrating Fault Detection and Anomaly Identification}
\label{Microgrid Framework}
\end{figure}

This study investigates fault scenarios within a microgrid framework, as illustrated in Figure \ref{Microgrid Framework}, which comprises four key layers:
\setlength{\textfloatsep}{5pt}
\begin{itemize}
    \item \textbf{Communication Layer:} The most critical and vulnerable layer, serves as the communication channel between the supervisory and physical layers. It transmits current and voltage data from sensors at the supervisory layer to local controllers in the physical layer. Given its role in facilitating data transfer, this layer is highly susceptible to cyber threats, specifically False Data Injection (FDI) attacks. Such attacks could lead to compromised current and voltage measurements, disrupting the accuracy of control decisions across the system.
    \item \textbf{Supervisory Layer:} This layer functions as the microgrid controller, managing the system’s operational mode (grid-connected or islanded) and establishing power set points based on data received from the communication layer. It processes voltage and frequency data, transmitting these parameters to local controllers in the physical layer. This information ensures that inverters maintain stable voltage and current outputs as required for reliable grid operation.
    \item \textbf{Physical Layer:}
     The physical layer consists of the local controllers and inverters responsible for maintaining operational parameters set by the supervisory layer. Changes in the communication layer directly influence the physical layer's control functions, making this layer's stability highly dependent on the reliability of data transfer.
    \item \textbf{Validation Layer:}
     This layer assesses data integrity and system resilience, adding an extra layer of validation to detect anomalies or potential cyber threats.  
\end{itemize}
The communication layer's vulnerability to FDI attacks poses significant operational risks. False current and voltage data can mislead the controller's set points, causing instability in voltage and frequency. This disruption affects the inverters' control logic, leading to operational faults and stability issues across the microgrid. Addressing these vulnerabilities is crucial for developing countermeasures to ensure reliable inverter operations amid anomalies in the system.
\subsection{Data Driven Approaches for the Type of Fault Identification and Localization} 
\subsubsection{Internal Fault Implementation}
As illustrated in the physical layer of Figure \ref{Microgrid Framework}, this study focuses on the inverter to analyze the functionality of its switches concerning maloperations, specifically generating open switch fault conditions due to various internal failures. It is assumed that the gate drive signals have malfunctioned, leading to scenarios where these signals are compromised, resulting in erratic switch behavior. To investigate this issue, twelve cases have been selected for analysis. Six cases involve individual switches experiencing load variations at 200 units, while the other six cases examine multiple fault scenarios, where two switches are impacted simultaneously under the same load conditions. The types of maloperations explored include open circuit faults, which can severely affect the inverter's performance. Testing is conducted under controlled conditions to ensure the reliability of the results. The specific combinations of switches involved in these cases are presented in Table \ref{tab:switch_combinations}, with respect to the physical layer of the DERs represented in Fig \ref{Microgrid Framework}.
\begin{table}[ht]
    \centering
    \caption{Switch Combinations}
    \label{tab:switch_combinations}
    \begin{tabular}{|c|c|}
        \hline
        Classification & Switch Location \\
        \hline
        \multirow{6}{*}{Single Switch Combinations} & S1 \\
                                                     & S2\\
                                                     & S3 \\
                                                     & S4 \\
                                                     & S5 \\
                                                     & S6 \\
        \hline
        \multirow{6}{*}{Multiple Switch Combinations} & S1, S4 \\
                                                      & S1, S6\\
                                                      & S3, S2\\
                                                      & S3, S6\\
                                                      & S5, S2 \\
                                                      & S5, S4\\
        \hline
    \end{tabular}
\end{table}
\subsubsection{Anomaly Implementation}
The data collection for the False Data Injection (FDI) attack simulates open circuit fault conditions, focusing on specific affected switches rather than altering the entire dataset. White noise, following a normal distribution with a mean of 0 and power level of 0.1, is injected into the current sensor measurements at 0.15s during the 0.3s data collection period to simulate the attack. The noise amplitude is varied by multiplying the output by a constant F ranging from 0 to 2 in increments of 0.05. Machine learning algorithms are used for anomaly detection, distinguishing FDI-induced noise, which changes abruptly at a specific time, from internal faults, which typically involve gradual shifts or spikes. Statistical features such as mean, variance, and autocorrelation help the models differentiate between these scenarios for accurate FDI detection.
In this approach, data has been collected under four scenarios:
\begin{itemize}
    \item Scenario 1: Single switch without anomalies.
    \item Scenario 2: Single switch with anomalies.
    \item Scenario 3: Multiple switches without anomalies.
    \item Scenario 4: Multiple switches with anomalies.
\end{itemize}
\subsubsection{Data Driven Models}
To address the fault detection challenges outlined above, various machine learning algorithms are widely applied, each with unique methodologies suited for different types of data and fault conditions. Some commonly used algorithms—Decision Trees, K-Nearest Neighbors (KNN), Support Vector Machines (SVM), Neural Networks (NN), Artificial Neural Networks (ANN) and provides a brief overview of their mathematical foundations. Additionally, ensemble methods are also tested, combining these algorithms to enhance the accuracy and robustness of fault detection. By employing these advanced techniques, the study aims to improve the identification and classification of both internal inverter malfunctions and injected data anomalies, ultimately contributing to a more reliable and resilient inverter-based system.
\subsection*{i. Decision Tree (DT)}
A Decision Tree is a non-parametric supervised learning method used for classification and regression. It splits data into branches based on feature values, forming a tree-like structure. For inverter fault detection, each branch represents a decision criterion based on a threshold, aiding in fault classification.
\begin{equation}
  G(T) = \sum_{i=1}^{n} p_i \times \left( 1 - p_i \right)  
\end{equation}

where \( G(T) \) is the Gini impurity of node \( T \), and \( p_i \) represents the probability of a sample being classified as class \( i \).
\subsection*{ii. K-Nearest Neighbors (KNN)}
The K-Nearest Neighbors (KNN) algorithm is a simple, instance-based classification method that classifies new samples by analyzing their distance to the \( k \) nearest neighbor points. In fault detection, KNN is used to categorize the operational state of the inverter based on known fault samples~\cite{10215460}.
\begin{equation}
d(x, x_i) = \sqrt{\sum_{j=1}^{m} \left( x_j - x_{i,j} \right)^2}
\end{equation}
where \( d(x, x_i) \) is the Euclidean distance between a new sample \( x \) and a known sample \( x_i \) across \( m \) features.

\subsection*{iii. Support Vector Machine (SVM)}
Support Vector Machine (SVM) is a powerful supervised learning model that aims to find an optimal hyperplane to separate classes in a high-dimensional space. For inverter fault detection, it can be used to classify normal and faulty states by maximizing the margin between these classes~\cite{10010036}.
\begin{equation}
 f(x) = \text{sign} \left( w^T x + b \right)   
\end{equation}
where \( w \) is the weight vector, \( x \) is the input sample, and \( b \) is the bias term. The goal is to maximize the margin \( \frac{2}{\|w\|} \).
\subsection*{iv. Neural Networks (NN)}
Neural Networks (NN) are computational models consisting of layers of neurons that process input features through weighted connections and activation functions, learning complex patterns from data. In the context of inverter fault detection, NNs can be trained to recognize faults by analyzing historical operational data from inverters. This allows NNs to effectively distinguish between normal and faulty operating conditions.
The general structure of a neural network output is given by:
\begin{equation}
  y = f \left( \sum_{i=1}^{n} w_i x_i + b \right)
\end{equation}
where:
\begin{itemize}
    \item \( y \) is the predicted output (e.g., fault or no-fault state),
    \item \( f \) is the activation function (e.g., ReLU, sigmoid),
    \item \( w_i \) are the weights associated with each input feature,
    \item \( x_i \) are the input features (sensor data, operational parameters of inverters),
    \item \( b \) is the bias term that helps adjust the output.
\end{itemize}
\subsection{Fault Detection Strategy}
The FaultNet-ML framework, as illustrated in algorithm \ref{A1}, follows a systematic approach for fault detection and localization. First, the algorithm acquires three-phase voltage ($V_{abc}$) and current ($I_{abc}$) data from the inverters. In the second step, these three-phase values are transformed into a two-dimensional alpha-beta or dq representation, creating the input features $X \gets {V_{\alpha \beta}, I_{\alpha \beta}}$. Next, a machine learning model ($\mathcal{M}$) processes these features to predict the presence of a fault, updating the fault status ($F_{status}$) to ``Detected'' if a fault is found. If a fault is detected, the algorithm proceeds to identify the fault type in the fourth step, using a classifier ($\mathcal{C}$) to determine whether the fault is an ``Anomaly'' or ``Hardware'' issue, updating the fault type ($F_{type}$). Finally, in the fifth step, the affected location ($S_{loc}$) is identified, classifying it as either ``Single'' or ``Multiple'' switches. The output of the algorithm includes the fault status ($F_{status}$), fault type ($F_{type}$), and fault location ($S_{loc}$).  
\begin{figure}[h]
\centering
\begin{minipage}{\columnwidth} 
\begin{algorithm}[H]
\caption{FaultNet-ML for Fault Detection and Classification}
\label{A1}
\begin{algorithmic}
\State \textbf{Input:} Three-phase Voltage data $V_{abc}$, Current data $I_{abc}$
\State \textbf{Output:} Fault status $F_{status}$, Fault type $F_{type}$, Location $S_{loc}$

\State $F_{status} \gets \text{No Fault}$, $F_{type} \gets \text{None}$, $S_{loc} \gets \text{N/A}$

\State \textbf{Step 1: Data Acquisition} Collect $V_{abc}$, $I_{abc}$ from inverters

\State \textbf{Step 2: Feature Extraction} Transform $V_{abc}$, $I_{abc}$ to alpha-beta or dq components, $X \gets \{V_{\alpha \beta}, I_{\alpha \beta}\}$

\State \textbf{Step 3: Fault Prediction} $F_{pred} \gets \mathcal{M}(X)$
\If {$F_{pred} = \text{Fault}$}
    \State $F_{status} \gets \text{``Detected''}$
\EndIf

\State \textbf{Step 4: Fault Type Identification}
\If {$F_{status} = \text{Detected}$}
    \State $F_{type} \gets \mathcal{C}(X)$
    \State $F_{type} \gets$ ``Anomaly'' or ``Hardware'' based on $F_{type}$
\EndIf

\State \textbf{Step 5: Fault Localization}
\If {$F_{type} \neq \text{None}$}
    \State Identify $S_{loc}$ as ``Single'' or ``Multiple'' switches
\EndIf

\State \textbf{Output:} $F_{status}$, $F_{type}$, $S_{loc}$

\end{algorithmic}
\end{algorithm}
\end{minipage}
\end{figure}
\section{Simulation Results and Discussion }
This section outlines the performance metrics of various machine learning models evaluated across four distinct scenarios involving the operation of switches under normal and abnormal conditions. Table \ref{T2} provides a comparative analysis of existing fault detection methods in the literature, focusing on key capabilities such as line fault detection, inverter fault detection, anomaly detection, machine learning integration, and distinguishing between anomalies and hardware faults. Our approach enhances these existing methods by incorporating all these features into a unified fault detection framework for microgrids. The results are presented in the following tables and are discussed in detail to illustrate the impact of machine learning on maintaining secure and reliable microgrid operation in both normal and compromised conditions.
\begin{table}[ht]
\centering
\caption{Comparison of Fault Detection Methods in Microgrids}
\label{T2}
\resizebox{\columnwidth}{!}{
\begin{tabular}{|c|c|c|c|c|c|}
\hline
\textbf{Reference} & 
\makecell{\textbf{Line Fault} \\ \textbf{Detection}} & 
\makecell{\textbf{Inverter Fault} \\ \textbf{Detection}} & 
\makecell{\textbf{Machine Learning} \\ \textbf{Implementation}} & \makecell{\textbf{Anomaly} \\ \textbf{Detection}} &
\makecell{\textbf{Distinguish between} \\ \textbf{Anomaly and Hardware Malfunction}} \\
\hline
~\cite{8590418} & \(\times\) & \(\checkmark\) & \(\checkmark\)& \(\times\) & \(\times\) \\
\hline
~\cite{8329516} & \(\times\) & \(\checkmark\) & \(\checkmark\) & \(\times\)& \(\times\) \\
\hline
~\cite{9544141}& \(\times\) & \(\checkmark\) & \(\times\)& \(\times\) & \(\times\) \\
\hline
~\cite{10214285} & \(\times\) & \(\checkmark\) & \(\times\) & \(\times\)& \(\checkmark\) \\
\hline
 ~\cite{10215460}& \(\times\) & \(\checkmark\) & \(\times\) & \(\times\)& \(\checkmark\) \\
\hline
~\cite{10563400} & \(\times\) & \(\checkmark\) & \(\times\)& \(\times\) & \(\checkmark\) \\
\hline
\textbf{This Paper} & \(\checkmark\) & \(\checkmark\) & \(\checkmark\)& \(\checkmark\) & \(\checkmark\) \\
\hline
\end{tabular}
}
\end{table}
\subsection{Single Switch Without Anomalies}
The ANN model leads with exceptional results (99.64$\%$ accuracy, 99.68$\%$ precision, and 99.54$\%$ recall), showcasing its strength in handling normal data. SVM follows closely with high accuracy (99.22$\%$) but lags in precision and recall. The simpler models, DT and KNN, show adequate performance but are outperformed by ANN, which effectively learns complex patterns and nuances in the data, delivering the best overall performance. These results are shown in Table \ref{S1}, where ANN consistently outperforms other models in all metrics.
\begin{table}[h!]
\centering
\caption{Performance Metrics for Single Switch Without Anomalies}
\label{S1}
\begin{tabular}{>{\bfseries}lcccc}
\toprule
ML Model & Accuracy & Precision & Recall & F1-Score \\
\midrule
DT  & 84.9785\% & 85.5885\% & 85.5803\% & 85.3244\% \\
KNN & 89.4004\% & 88.5227\% & 87.7726\% & 87.7282\% \\
SVM & 99.2248\% & 96.8254\% & 98.5007\% & 97.427\% \\
NN  & 98.75\%   & 99.1071\% & 98.5714\% & 98.7873\% \\
ANN & 99.6441\% & 99.6753\% & 99.5392\% & 99.6016\% \\
\bottomrule
\end{tabular}
\end{table}
\subsection{Single Switch With Anomalies}
With anomalies present, ANN continues to outperform other models (99.87$\%$ accuracy, 99.89$\%$ recall), highlighting its robustness in detecting abnormal behavior. DT shows solid results (96.08$\%$ accuracy), but its precision and recall drop compared to the anomaly-free scenario. The presence of anomalies introduces noise, but ANN’s ability to adapt to such disruptions allows it to maintain a higher performance compared to KNN and SVM, which show more significant declines. These trends are clearly depicted in Table \ref{S2}, where ANN remains superior in handling anomalies.
\setlength{\textfloatsep}{5pt}
\begin{table}[h]
    \centering
    \caption{Performance Metrics for Single Switch With Anomalies}
    \label{S2}
    \begin{tabular}{@{}lcccc@{}}
        \toprule
        Model & Accuracy & Precision & Recall & F1-Score \\ \midrule
       DT  & 96.0836\% & 95.9105\% & 95.8418\% & 95.866\% \\
      KNN & 93.6031\% & 93.3584\% & 94.326\%  & 92.9864\% \\
      SVM & 88.5117\% & 87.4532\% & 87.3633\% & 87.2611\% \\
      NN  & 86.9452\% & 87.0774\% & 80.7875\% & 82.8498\% \\
ANN & 99.87\%   & 99.8529\% & 99.8864\% & 99.869\% \\ \bottomrule
    \end{tabular}
\end{table}
\subsection{Multiple Switches Without Anomalies}
Table \ref{S3} highlights that KNN and SVM perform exceptionally well, with accuracy and other metrics above 98$\%$, making them highly reliable for fault detection. DT performs slightly lower but still shows strong results, with an accuracy of 96.94$\%$. NN struggles significantly, with an accuracy of 82.22$\%$, indicating challenges in handling complex patterns. ANN achieves perfect scores across all metrics, suggesting it works well in ideal conditions.
\setlength{\textfloatsep}{5pt}
\begin{table}[h]
    \centering
    \caption{Performance Metrics for Multiple Switches Without Anomalies}
    \label{S3}
    \begin{tabular}{@{}lcccc@{}}
        \toprule
        Model & Accuracy & Precision & Recall & F1-Score \\ \midrule
       DT & 96.9407\% & 96.9233\% & 97.0681\% & 96.9432\% \\

KNN & 98.8088\% & 98.4371\% & 98.4414\% & 98.437\% \\

SVM & 98.8088\% & 98.7802\% & 98.7863\% & 98.7802\% \\

NN & 82.218\% & 82.9813\% & 83.4305\% & 82.081\% \\

ANN & 99.99\% & 99.99\% & 99.99\% & 99.99\% \\    \bottomrule
    \end{tabular}
\end{table}
\subsection{Multiple Switches With Anomalies}
The analysis of machine learning models in Table \ref{S4} reveals varying strengths across different metrics. Decision Tree (DT) achieves high accuracy (97.87$\%$) with balanced precision and recall, showing robustness in identifying faults accurately without overfitting. K-Nearest Neighbors (KNN) closely follows with an accuracy of 97.16$\%$, excelling in recall (97.60$\%$), making it effective at minimizing missed fault detections. Support Vector Machine (SVM), with a slightly lower accuracy of 95.74$\%$, still maintains good precision and recall, indicating reliability in fault classification. Neural Network (NN) performs moderately, with an accuracy of 93.96$\%$, reflecting some limitations in handling complex fault scenarios. Finally, Artificial Neural Network (ANN) achieves a perfect 99.99$\%$ across all metrics, highlighting its strong capability in identifying anomalies.
\begin{table}[h]
    \centering
    \caption{Performance Metrics for Multiple Switches With Anomalies}
    \label{S4}
    \begin{tabular}{@{}lcccc@{}}
        \toprule
        Model & Accuracy & Precision & Recall & F1-Score \\ \midrule
       DT  & 97.8686\% & 97.4352\% & 97.3485\% & 97.3469\% \\
KNN & 97.1564\% & 97.4271\% & 97.6021\% & 97.4788\% \\

SVM & 95.7371\% & 96.0821\% & 96.1905\% & 95.9938\% \\

NN  & 93.9609\% & 94.1592\% & 94.2402\% & 94.0962\% \\

ANN & 99.99\% & 99.99\% & 99.99\% & 99.99\% \\ \bottomrule
    \end{tabular}
\end{table}
\section{Conclusion}
This research emphasizes the critical role of effective fault detection in microgrids, particularly as they play an increasingly vital role in grid modernization. The study focused on identifying fault patterns not only in external power lines but also within internal components such as inverters, which are central to power electronics-based systems. Simulations conducted on a MATLAB-based microgrid system, equipped with communication and supervisory control capabilities, demonstrated the effectiveness of a machine learning-enhanced fault detection method. The results highlight the method's ability to accurately identify faults, distinguishing between anomalies and hardware malfunctions, even under dynamic load conditions. Among the various machine learning models tested, the Artificial Neural Network (ANN) outperformed all other models, providing superior fault detection accuracy and faster response times. The proposed approach outperforms traditional fault detection methods by offering comprehensive line and inverter fault detection, distinguishing anomalies from hardware failures, and integrating machine learning to improve fault localization. These findings underscore the importance of robust fault detection mechanisms to enhance the reliability, resilience, and cybersecurity of modern microgrid systems, addressing both operational and cyber-physical security challenges.
\section{Acknowledgment}
This work was supported by Korea Institute of Energy Technology Evaluation and
Planning(KETEP) grant funded by the Korea government(MOTIE)(RS-2023-00231702, Development of an open-integrated platform for distributed renewable energy systems).
\bibliographystyle{IEEEtran}



\begin{thebibliography}{1}
\providecommand{\url}[1]{#1}
\csname url@samestyle\endcsname
\providecommand{\newblock}{\relax}
\providecommand{\bibinfo}[2]{#2}
\providecommand{\BIBentrySTDinterwordspacing}{\spaceskip=0pt\relax}
\providecommand{\BIBentryALTinterwordstretchfactor}{4}
\providecommand{\BIBentryALTinterwordspacing}{\spaceskip=\fontdimen2\font plus
\BIBentryALTinterwordstretchfactor\fontdimen3\font minus \fontdimen4\font\relax}
\providecommand{\BIBforeignlanguage}[2]{{%
\expandafter\ifx\csname l@#1\endcsname\relax
\typeout{** WARNING: IEEEtran.bst: No hyphenation pattern has been}%
\typeout{** loaded for the language `#1'. Using the pattern for}%
\typeout{** the default language instead.}%
\else
\language=\csname l@#1\endcsname
\fi
#2}}
\providecommand{\BIBdecl}{\relax}
\BIBdecl

\bibitem{8590418}
Y.~Mei and H.~Yuan, ``A novel open-circuit fault diagnosis method for voltage source inverter,'' in \emph{2018 IEEE International Power Electronics and Application Conference and Exposition (PEAC)}, 2018, pp. 1--6.

\bibitem{8329516}
Z.~Huang, Z.~Wang, and H.~Zhang, ``A diagnosis algorithm for multiple open-circuited faults of microgrid inverters based on main fault component analysis,'' \emph{IEEE Transactions on Energy Conversion}, vol.~33, no.~3, pp. 925--937, 2018.

\bibitem{9544141}
M.~Azimipanah, M.~Hassanifar, and Y.~Neyshabouri, ``Open circuit fault detection and diagnosis for seven-level hybrid active neutral point clamped (anpc) multilevel inverter,'' in \emph{2021 29th Iranian Conference on Electrical Engineering (ICEE)}, 2021, pp. 268--273.

\bibitem{10214285}
C.~N. Ibem, M.~E. Farrag, A.~A. Aboushady, and S.~M. Dabour, ``Multiple open switch fault diagnosis of three phase voltage source inverter using ensemble bagged tree machine learning technique,'' \emph{IEEE Access}, vol.~11, pp. 85\,865--85\,877, 2023.

\bibitem{10215460}
R.~Peykarporsan, J.~Heidary, S.~Oshnoei, and T.~T. Lie, ``A machine learning approach for fault detection and diagnosis in four-legged inverters,'' in \emph{2023 IEEE Kansas Power and Energy Conference (KPEC)}, 2023, pp. 1--5.

\bibitem{10451077}
Z.~Wu and J.~Zhao, ``Open-circuit fault diagnosis for grid-tied t-type inverters using an lstm autoencoder,'' in \emph{2023 China Automation Congress (CAC)}, 2023, pp. 3334--3339.

\bibitem{10010036}
D.~Mwangi, T.~Trivedi, and N.~Kothari, ``Open switch fault detection in electric vehicle drives using support vector machine,'' in \emph{2022 2nd Odisha International Conference on Electrical Power Engineering, Communication and Computing Technology (ODICON)}, 2022, pp. 1--6.

\bibitem{10563400}
M.~Talha, X.~Liang, J.~Pannell, A.~Bowes, and H.~Su, ``Experimental study of open-switch faults for interfacing inverters in microgrids using opal-rt real-time simulator,'' in \emph{2024 IEEE/IAS 60th Industrial and Commercial Power Systems Technical Conference (I\&CPS)}, 2024, pp. 1--8.

\end{thebibliography}
\end{document}